\documentclass[reprint,amsmath,amssymb,floatfix,superscriptaddress]{revtex4-2}
\usepackage{graphicx}
\usepackage{hyperref}

\newcommand*\mean[1]{\langle #1 \rangle}
\def\al{\hat{\alpha}}

\begin{document}
\title{Observation of quasiparticle pair-production and quantum entanglement in atomic quantum gases quenched to an attractive interaction}
\author{Cheng-An Chen}
\affiliation{Department of Physics and Astronomy, Purdue University, West Lafayette, IN 47907, USA}
\author{Sergei Khlebnikov}
\affiliation{Department of Physics and Astronomy, Purdue University, West Lafayette, IN 47907, USA}
\affiliation{Purdue Quantum Science and Engineering Institute, Purdue University, West Lafayette, IN 47907, USA}
\author{Chen-Lung Hung}
\email{clhung@purdue.edu}
\affiliation{Department of Physics and Astronomy, Purdue University, West Lafayette, IN 47907, USA}
\affiliation{Purdue Quantum Science and Engineering Institute, Purdue University, West Lafayette, IN 47907, USA}
\date{\today}

\begin{abstract}
We report observation of quasiparticle pair-production and characterize quantum entanglement created by a modulational instability in an atomic superfluid. By quenching the atomic interaction to attractive and then back to weakly repulsive, we produce correlated quasiparticles and monitor their evolution in a superfluid through evaluating the in situ density noise power spectrum, which essentially measures a `homodyne' interference between ground state atoms and quasiparticles of opposite momenta. We observe large amplitude growth in the power spectrum and subsequent coherent oscillations in a wide spatial frequency band within our resolution limit, demonstrating coherent quasiparticle generation and evolution. The spectrum is observed to oscillate below a quantum limit set by the Peres-Horodecki separability criterion of continuous-variable states, thereby confirming quantum entanglement between interaction quench-induced quasiparticles.
\end{abstract}
\maketitle

Coherent pair-production processes are enabling mechanisms for entanglement generation in continuous variable states \cite{keller1997theory,pan2012multiphoton}. In many-body systems, quasiparticle pair-production presents an interesting case, as interaction creates entanglement shared among collectively excited interacting particles. Entanglement distribution through quasiparticle propagation is a direct manifestation of transport property in a quantum many-body system \cite{eisert2015quantum,finazzi2014entangled}. Controlling quasiparticle pair-production and detecting entanglement evolution thus opens a door to probing quantum many-body dynamics, enabling fundamental studies such as information propagation \cite{jurcevic2014quasiparticle,cheneau2012light}, entanglement entropy evolution \cite{calabrese2004entanglement}, many-body thermalization \cite{abanin2019colloquium}, as well as Hawking radiation of quasiparticles and thermodynamics of an analogue black hole \cite{steinhauer2016observation,de2019observation,hu2019quantum}.

In atomic quantum gases, coherent quasiparticle pair-production can be stimulated through an interaction quench, which results in a rapid change of quasiparticle dispersion relation that can project collective excitations, from either existing thermal or quantum populations, into a superposition of correlated quasiparticle pairs \cite{jaskula2012acoustic,hung2013cosmology,schemmer2018monitoring}. This has led to prior observation of Sakharov oscillations in a quenched atomic superfluid \cite{hung2013cosmology,ranccon2013quench}. An intriguing case occurs when the atomic interaction is quenched to an attractive value, where quasiparticles become unstable. In this case, the early time dynamics is governed by a modulational instability (MI), which stimulates exponential growth of density waves. This growth leads eventually to wave fragmentation and soliton formation \cite{strecker2002formation}. Although consequences of MI have been observed in a number of recent quantum gas experiments \cite{nguyen2017formation,everitt2017observation,mevznarvsivc2019cesium,sanz2019interaction,chen2020observation}, the early-time evolution itself has only been recently studied \cite{chen2020observation}, and the direct verification of coherent quasiparticle pair-production has remained an open question.

\begin{figure}[b]
\includegraphics[width=1\columnwidth]{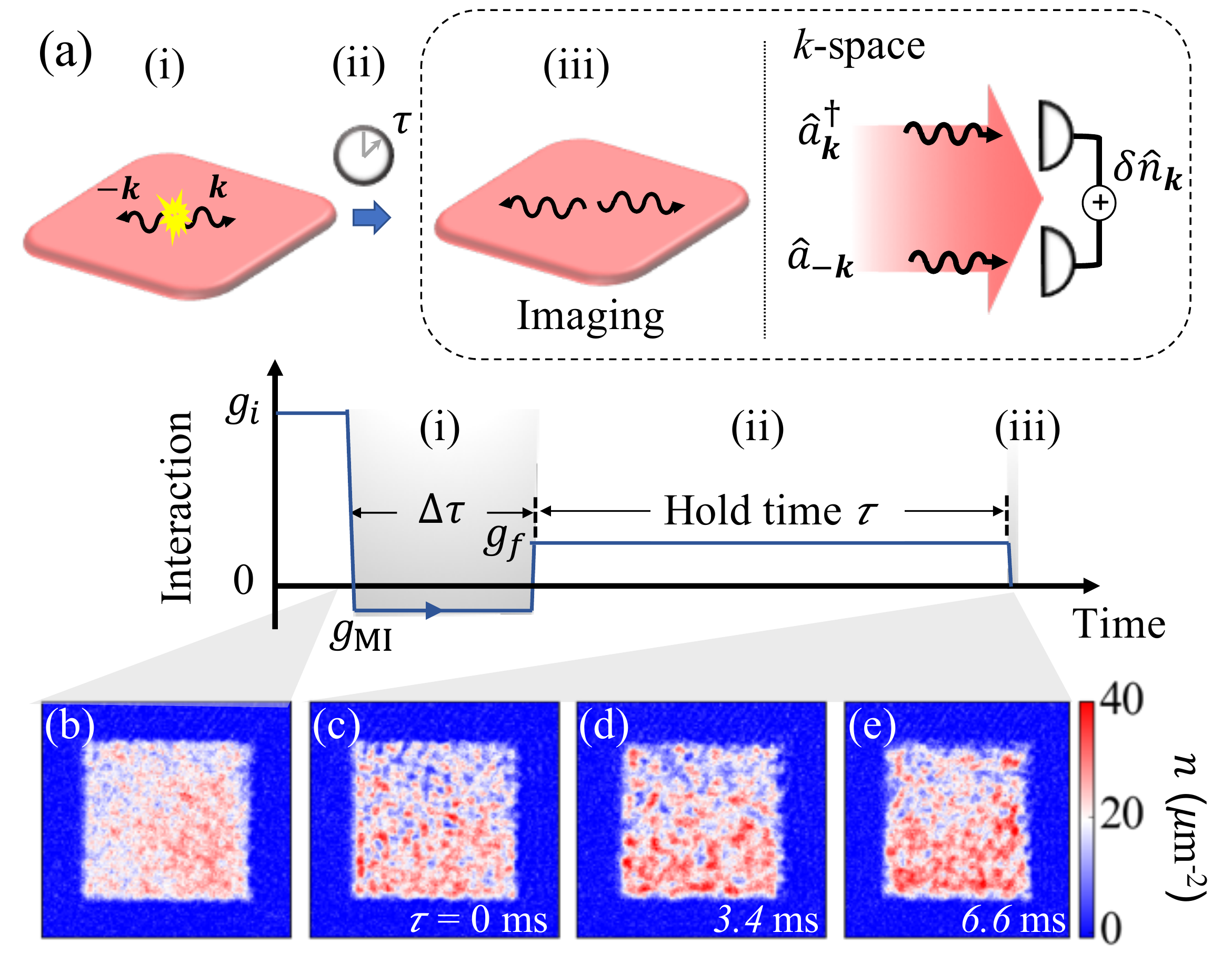}
\caption{Experiment scheme for quasiparticle pair-production and detection. (a) A homogeneous 2D superfluid (red square) undergoes an interaction quench protocol from (i) $g=g_i>0$ to $g_\mathrm{MI} < 0$ for broadband generation of quasiparticle pairs of opposite momenta (illustrated by black curvy arrows) for a time duration $\Delta\tau$; (ii) A second interaction quench to $g=g_f>0$ allows quasiparticles to evolve as phonons for a variable hold time $\tau$; (iii) In situ density noise in spatial frequency domain, $\delta n_\mathbf{k}$, is essentially a `homodyne' measurement of excitations in opposite momentum states interfering with ground state atoms. (b-e) Single-shot density images taken prior to (b) or after the interaction quench (c-e) and held for the indicated time $\tau$. Image size: 77$\times$77~$\mu$m$^2$.}
\label{fig1}
\end{figure}

In this letter, we demonstrate coherent quasiparticle pair-production by inducing MI in a homogeneous 2D quantum gas quenched to an attractive interaction. Subsequent quasiparticle evolution is monitored through quenching the interaction back to a positive value (Fig.~\ref{fig1}). Through in situ imaging, we analyze the dynamics of density observables by a method analogous to the well-established homodyne detection technique in quantum optics \cite{furusawa1998unconditional,braunstein1998teleportation,lvovsky2009continuous} and confirm non-classical correlations, that is, quantum entanglement in interaction quench-produced quasiparticle pairs.

Our analyses are based on the time evolution of in situ density noise, which is a manifestation of interference between quasiparticle excitations and the ground state atoms that serve as a coherent local oscillator \cite{ferris2008detection}. In Fourier space, the density noise operator can be written as $\delta \hat{n}_\mathbf{k} \approx \sqrt{N} (\hat{a}^\dag_\mathbf{k} + \hat{a}_\mathbf{-k})$, where $N \gg1$ is the total atom number nearly all accounted for by the ground state atoms, and $\hat{a}^{(\dag)}_\mathbf{\pm k}$ are the annihilation (creation) operators for $\pm\mathbf{k}$ single-particle momentum eigenstates. They are related to quasiparticle operators $\al^\dag_\mathbf{\pm k}$ by the Bogoliubov transformation.

\begin{figure}[t]
\includegraphics[width=.85\columnwidth]{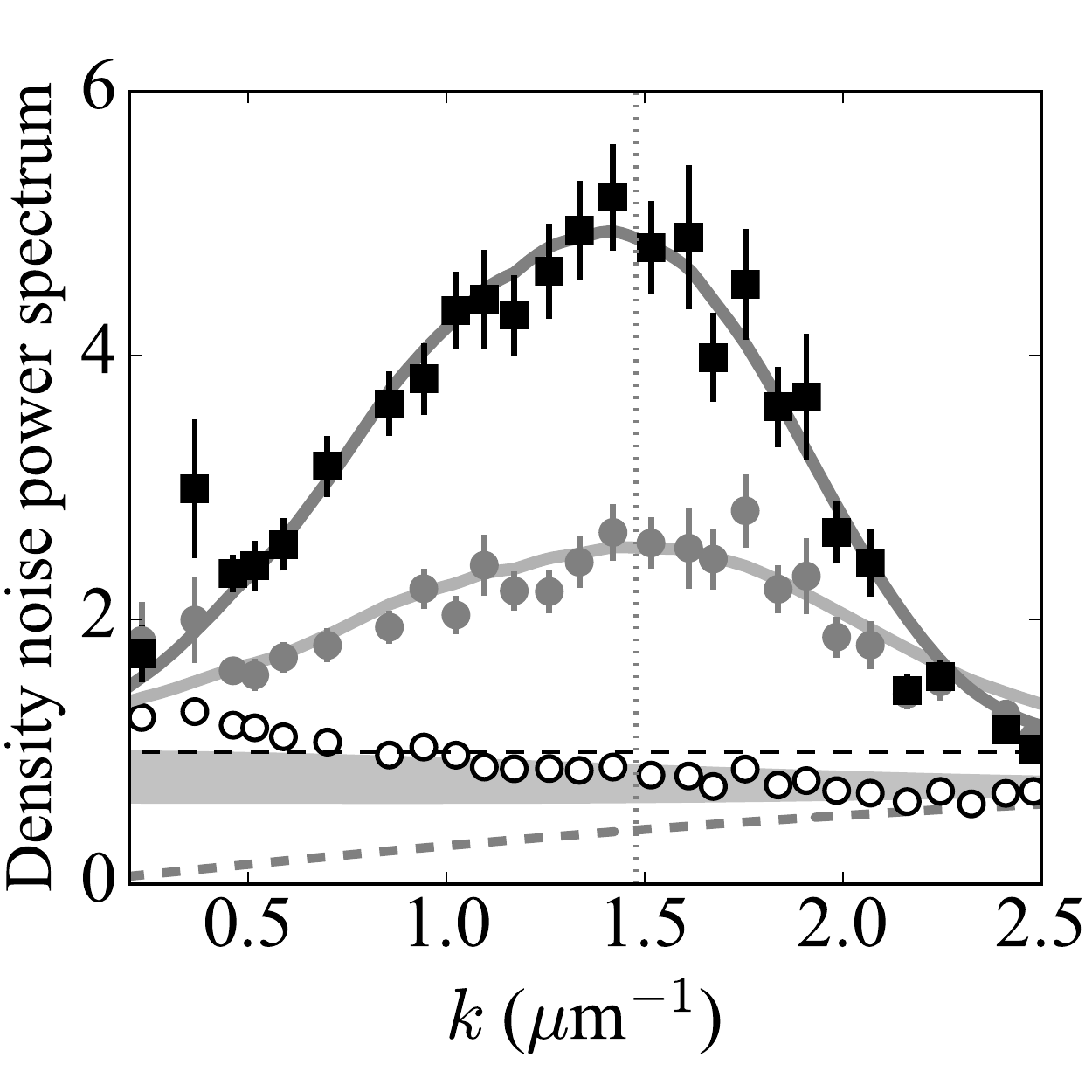}
\caption{Growth of density noise during the MI period. Density noise power spectra measured before, $S_0(k)$ (open circles), and right after the MI period, $S(k,0)$, with $\Delta\tau \approx 1~$ms (gray circles) and 2~ms (black squares), respectively. Horizontal dashed line marks the atomic shot-noise level. Gray band represents calculated initial phonon spectrum assuming equilibrium temperature $T=8\pm2$~nK. Dashed curve shows the squeezing parameter $C_k$ at $g=g_i\approx 0.127$. Solid curves are theory fits to data; see text \cite{SM}. Vertical dotted line marks the wavenumber $k_c$, below (above) which quasiparticles are expected to be unstable (stable) at $g=g_\mathrm{MI}\approx-0.026$.}
\label{fig2}
\end{figure}

To characterize non-classical correlation between quasiparticles, we study the density noise power spectrum $S(\mathbf{k}) = \mean{|\delta n_\mathbf{k} |^2}/N$, where $\mean{\cdots}$ denotes ensemble averaging. Within our resolution limit ($|\mathbf{k}|\lesssim 2.6/\mu$m), the power spectrum conveniently measures the variances of two-mode quadrature of quasiparticles of all $\pm \mathbf{k}$ momentum pairs, $\hat{x}_{\mathbf{k}}+\hat{x}_{-\mathbf{k}}$ and $\hat{p}_{\mathbf{k}}-\hat{p}_{-\mathbf{k}}$, where $\hat{x}_{\mathbf{k}} = (\al^\dag_{\mathbf{k}} + \al_{ \mathbf{k}})/\sqrt{2}$ and $\hat{p}_{ \mathbf{k}} = i( \al^\dag_{\mathbf{k}} - \al_{\mathbf{k}})/\sqrt{2}$ \cite{SM}. Since pair-production should be isotropic in our quantum gas samples, in the following we discuss azimuthally averaged spectrum $S(k)$, and use $\pm k$ to denote opposite momenta.

It suffices to prove inseparability or quantum entanglement in quasiparticles of opposite momenta by violating the Peres-Horodecki separability criterion, which -- in the continuous-variable version \cite{duan2000inseparability,simon2000peres} and adapted to our case -- reads $\left[\langle (\hat{x}_{k} + \hat{x}_{-k})^2\rangle + \langle(\hat{p}_{k} - \hat{p}_{-k})^2 \rangle\right] \geq 2$.
In terms of the density power spectrum, this becomes \cite{SM}
\begin{equation}
S(k)= \frac{C_k}{2} \left[\langle (\hat{x}_{k} + \hat{x}_{-k})^2\rangle + \langle(\hat{p}_{k} - \hat{p}_{-k})^2 \rangle\right] \geq C_k, \label{skrelation}
\end{equation}
where $C_k= \epsilon_k/\epsilon(k,g)$ is a squeezing parameter determined by the ratio between the single-particle energy $\epsilon_k$ and the phonon dispersion relation $\epsilon(k,g)=\sqrt{\epsilon_k^2 + 2\frac{\hbar^2}{m}\bar{n} g \epsilon_k}$, $\bar{n}$ is the mean density, $g$ is the interaction at the time of the measurement, $m$ is the atomic mass, and $\hbar$ is the reduced Planck constant. For non-interacting particles, we would have $C_k=1$, representing the limit of atomic shot-noise. Proving quasiparticle (phonon) entanglement in a superfluid ($g > 0$), on the other hand, requires a lower quantum limit $C_k<1$.

Physical interpretation of Eq.~(\ref{skrelation}) becomes clearer when we express the density noise power spectrum as an observable for superfluid phonon populations and time-dependent pair-correlation signal (at $g>0$) \cite{SM,robertson2017controlling,robertson2017assessing}:
\begin{equation}
S(k,\tau) = C_k \left[1+\bar{N}_k + \Delta N_k \cos \phi_k(\tau) \right],\label{phonon}
\end{equation}
where $\bar{N}_k=\mean{\hat{\alpha}^\dag_k\hat{\alpha}_k}+\mean{\hat{\alpha}^\dag_{-k}\hat{\alpha}_{-k}}$ is the mean phonon number in $\pm k$ modes, while $ \Delta N_k$ and $\phi_k (\tau)$ are related to the amplitude and argument of the pair-correlation observable $\mean{\hat{\alpha}^\dag_k\hat{\alpha}^\dag_{-k}}$ (and $\mean{\hat{\alpha}_k\hat{\alpha}_{-k}}$), respectively. The pair-correlation observable oscillates as time evolves due to a dynamical phase factor $\phi_k(\tau)$ accumulating between phonons of opposite momenta. Violating Eq.~(\ref{skrelation}) is equivalent to having $\Delta N_k>\bar{N}_k$, resulting in maximal two-mode squeezing $S(k)/ C_k <1$ at $\phi_k \approx (2l+1)\pi$ and anti-squeezing $S(k)/C_k > 1$ at $\phi_k \approx 2l\pi$ ($l$ is an integer) -- a key entanglement signature that we demonstrate in this letter.

To carry out the experiment, we prepare uniform superfluid samples formed by $N\approx 4.9\times 10^4$ nearly pure Bose-condensed cesium atoms loaded inside a quasi-2D box potential, which compresses all atoms in the harmonic ground state along the imaging ($z$-) direction \cite{chen2020observation} with $l_z=184~$nm being the harmonic oscillator length. A time-of-flight measurement estimates the sample temperature $T\lesssim 8~$nK. Mean atomic surface density $\bar{n} \approx 21/\mu$m$^2$ is approximately uniform within a horizontal box size of $\approx 48\times 48~\mu$m$^2$. The interaction strength of the quasi-2D gas $g =\sqrt{8\pi}a/l_z$ is controlled by the s-wave scattering length $a$ via a magnetic Feshbach resonance \cite{chin2010feshbach}, giving an initial interaction strength $g=g_i \approx 0.127$. An uncertainty in $g$ ($\delta g \approx \pm 0.0006$) is primarily contributed by the uncertainty in the magnetic field at the scattering length zero-crossing \cite{chen2020observation}.

Our interaction quench protocol is illustrated in Fig.~\ref{fig1}(a). A MI period is initiated by quenching the atomic interaction (within $0.8~$ms) to a negative value $g_\mathrm{MI}\approx-0.026$ and holding for a short time $\Delta \tau \approx 1\sim2~$ms. To terminate MI, we quench the atomic interaction back to a small positive value $g_f\approx0.007$, allowing quasiparticles to evolve as phonons in a stable superfluid for another variable time $\tau$ before we perform in situ absorption imaging. Figures~\ref{fig1}(b-e) show sample images measured before and after we initiate the quench protocol. We evaluate $\delta n_\mathbf{k}$ for each sample through Fourier analysis \cite{hung2011extracting} and obtain their density noise power spectra. Typically around 50 experiment repetitions are analyzed for each hold time $\tau$. Each power spectrum has been carefully calibrated with respect to the atomic shot-noise measured from high temperature normal gases \cite{hung2011extracting,SM}.

To confirm quasipaticle generation during the MI period, in Fig.~\ref{fig2} we compare the density noise power spectra measured before and immediately after the MI period, that is, for hold time $\tau=0$. Before MI, the initial spectrum $S_0(k)$ is mostly below the atomic shot-noise due to low temperature $T\lesssim8$~nK and small initial squeezing parameter $C_k<1$. Excessive noise in $k\lesssim 0.75/\mu$m may be due to technical heating in the box potential. After the MI time period $\Delta \tau$, we indeed find significant increase in the density noise $S(k,0)>1$ at spatial frequencies extending over two quasiparticle regimes: $k \lesssim k_c = 2\sqrt{\bar{n} |g_\mathrm{MI}|}\approx 1.5/\mu$m in the instability band showing hyperbolic growth due to a purely imaginary dispersion relation $\epsilon(k,g_\mathrm{MI})$ \cite{chen2020observation}, and $k\gtrsim k_c$ in the stable regime showing quasiparticle
production due to variation of the interaction strength \cite{hung2013cosmology}.

\begin{figure}[t]
\includegraphics[width=1\columnwidth]{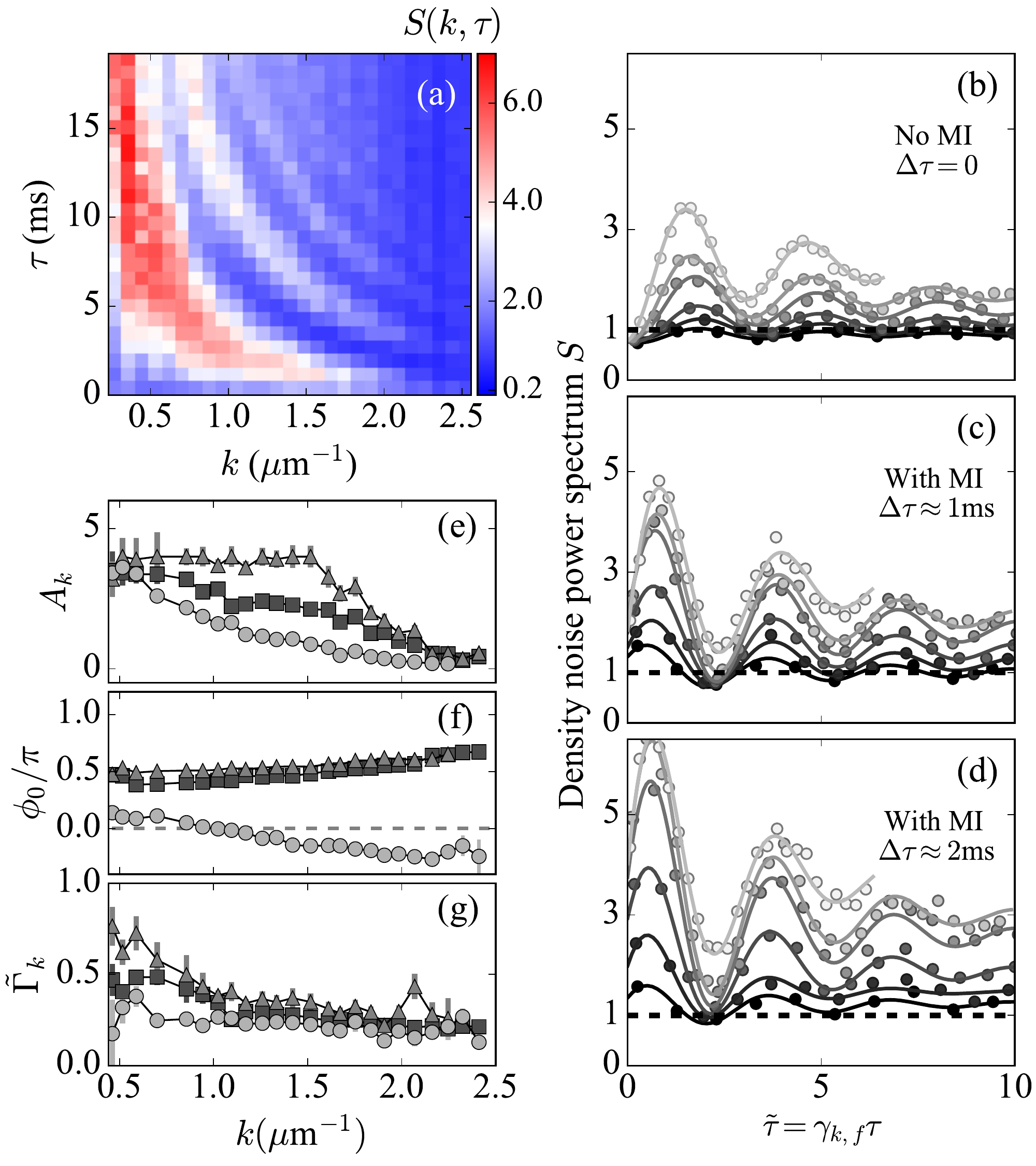}
\caption{Coherent oscillations in the density noise power spectrum. (a) Full evolution of the power spectrum $S(k,\tau)$ with $\Delta\tau \approx 1~$ms, showing coherent oscillations in time and $k$-space. (b-d) Synchronized oscillations of $S(k,\tilde{\tau})$ plotted in the rescaled time unit $\tilde{\tau}=\gamma_{k,f}\tau$ for various $k\approx (1, 1.3, 1.6, 1.8, 2.1, 2.2)/\mu$m (Gray circles from bright to dark). Horizontal dashed lines mark the atomic shot-noise limit. Solid lines are sinusoidal fits. Fitted amplitude $A_k$, phase offset $\phi_0$, and decay rate $\tilde{\Gamma}_k$ from samples with $\Delta \tau\approx 0$~ms (filled circles), 1~ms (filled squares), and 2~ms (filled triangles) are plotted in (e-g), respectively.}
\label{fig3}
\end{figure}

Our measured spectra can be well-captured by a model $S(k,0) = e^{-\Gamma_k\Delta \tau}S_\mathrm{coh}(k) + S_\mathrm{inc}(k) $, which incorporates the Bogoliubov theory and quasiparticle dissipation through coupling to a single-particle bath \cite{SM}. The first term is a damped coherent signal $S_\mathrm{coh}(k) = S_0(k)[ 1 + \frac{\epsilon(k,g_i)^2-\epsilon(k,g_\mathrm{MI})^2}{\epsilon(k,g_\mathrm{MI})^2}\sin^2\frac{\epsilon(k,g_\mathrm{MI})\Delta\tau}{\hbar}]$ \cite{chen2020observation,hung2013cosmology}, while the second term is referred to as an incoherent signal $S_\mathrm{inc}(k) = \frac{1}{2}\{\eta_{-}\frac{\Gamma_k^2}{\Gamma_k^2 + 4\epsilon(k,g_\mathrm{MI})^2/\hbar^2}[ 1-e^{-\Gamma_k\Delta\tau}(\cos\frac{2\epsilon(k,g_\mathrm{MI})\Delta\tau}{\hbar} - \frac{2\epsilon(k,g_\mathrm{MI})}{\hbar\Gamma_k}\sin \frac{2\epsilon(k,g_\mathrm{MI})\Delta \tau}{\hbar})] + \eta_{+}(1-e^{-\Gamma_k\Delta\tau})\}$, where $\eta_{\pm}=1 \pm \epsilon_k^2/\epsilon(k,g_\mathrm{MI})^2$. Our theory fits (solid curves in Fig.~\ref{fig2}) suggest a $k$-dependent dissipation rate $\Gamma_k \sim 0.5 \epsilon_k/\hbar$ \cite{SM}, of the same order of magnitude as the decay rate extracted from the subsequent time-evolution measurements at $g=g_f$ (Fig.~\ref{fig3}).

To demonstrate phase coherence and pair-correlation in these interaction quench-induced quasiparticle pairs, we plot the complete time and momentum dependence of the density noise power spectrum $S(k,\tau)$, as shown in Fig.~\ref{fig3}(a). Here, oscillatory behavior is clearly visible over the entire spectrum. The oscillations are a manifestation of the interference between coherent quasiparticles of opposite momenta $\pm k$, as suggested by Eq.~(\ref{phonon}), with the relative phase winding up in time as $\phi_k(\tau)=2\gamma_{k,f} \tau+\phi_0$, where $\gamma_{k,f}=\epsilon(k,g_f)/\hbar$ is the expected Bogoliubov phonon frequency and $\phi_0$ is an initial phase difference. In Fig.~\ref{fig3}(b-d), we plot $S(k,\tilde{\tau})$ in the rescaled time $\tilde{\tau}=\gamma_{k,f}\tau$ and confirm that all spectra oscillate synchronously with a time period $\approx \pi$, thus validating the phonon interference picture. For comparison, we also plot the evolution of samples with a direct interaction quench from $g_i$ to $g_f$ without an MI period ($\Delta \tau=0$). Oscillations in $S(k,\tilde{\tau})$ can also be observed, albeit with smaller amplitudes and phase offsets $\phi_0 \approx 0$, as these oscillations result solely from the interference of in-phase quasiparticle projections from suddenly decreasing the Bogoliubov energy \cite{hung2013cosmology}. In either case, with or without MI, we observe that phase coherence is lost in a few cycles and the density noise spectra reach new steady-state values.

To quantify phase coherence and dissipation at final $g=g_f$, we perform simple sinusoidal fits $S(k,\tilde{\tau})= S_{f}-S_o e^{-\tilde{\Gamma}_k \tilde{\tau}} - A_k e^{-\tilde{\Gamma}_k \tilde{\tau}}\cos(2\tilde{\tau} + \phi_0)$ to the data to extract ($A_k$, $\phi_0$, $\tilde{\Gamma}_k$), as shown in Fig.~\ref{fig3}(e-g) (the steady-state values $S_f$ and $S_o$ are not shown). The larger oscillation amplitudes $A_k$ found in samples with $\Delta\tau\approx 1~$ms and $2~$ms show that MI-induced quasiparticles are highly phase coherent. This can also be seen in the non-zero phase offset $\phi_0\gtrsim \pi/2$ at $k\gtrsim 0.5/\mu$m in Fig.~\ref{fig3}(f), which is coherently accumulated during the MI period. Furthermore, in Fig.~\ref{fig3}(g), we observe a nearly constant decay rate $\tilde{\Gamma}_k \approx 0.31\pm0.08$ at $k\gtrsim0.8~/\mu$m for these MI-induced oscillations. This is close to the decay rate $\tilde{\Gamma}_k \approx 0.22\pm0.04$ in samples without an MI period ($\Delta\tau=0$), suggesting that the short MI dynamics does not heat up the sample significantly to increase the phonon dissipation rate.

\begin{figure}[t]
\includegraphics[width=1\columnwidth]{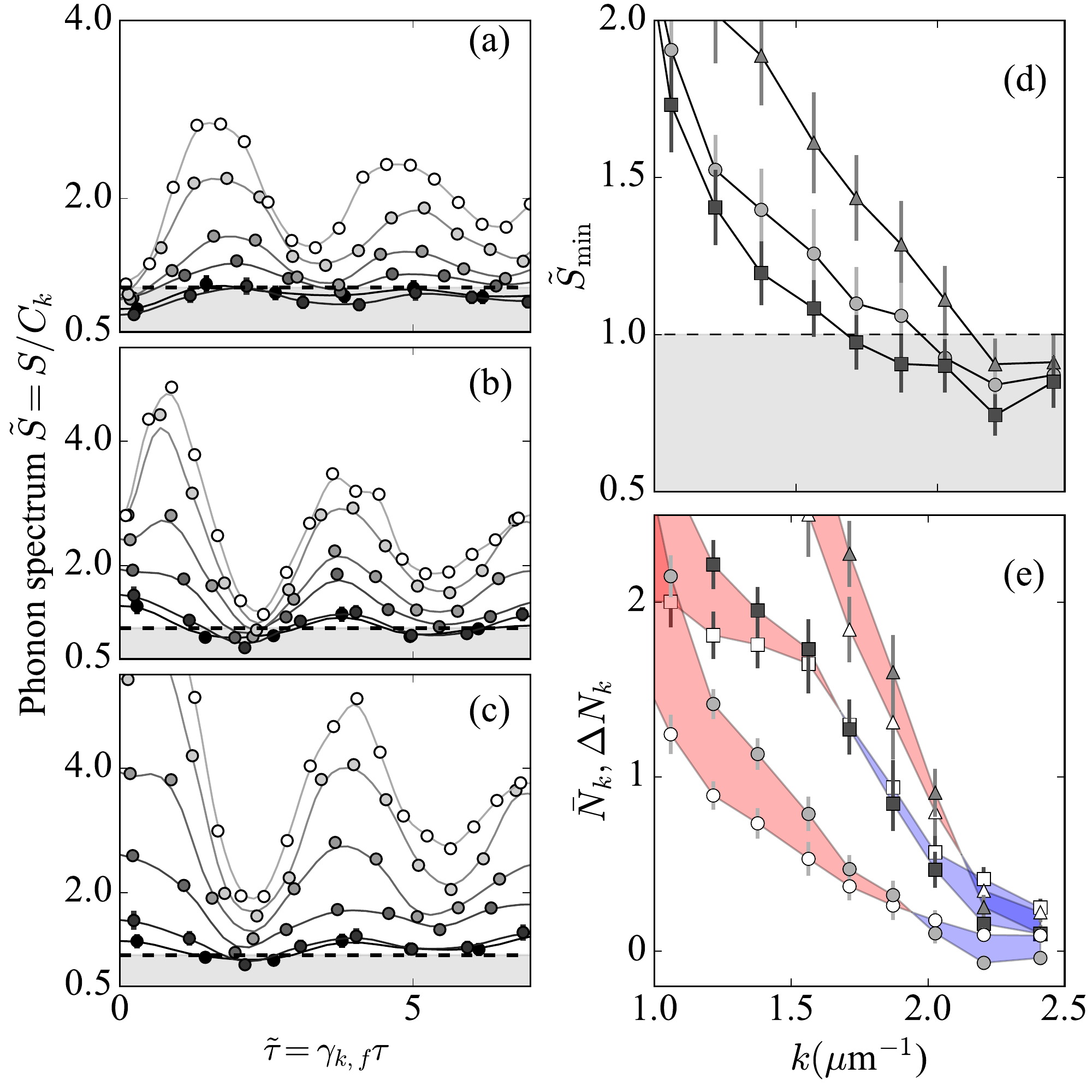}
\caption{Testing two-mode squeezing and quantum entanglement in the phonon basis. (a-c) Rescaled phonon spectrum $\tilde{S}(k,\tilde{\tau})$ for $k\approx ( 1.3, 1.6, 1.8, 2.1, 2.2, 2.4)/\mu$m (filled circles from bright to dark), evaluated using data as shown in Figs.~\ref{fig3}(b-d). Solid curves are guides to the eye. (d) First minima $\tilde{S}_\mathrm{min}$ in the phonon spectra of various wavenumber $k$, at $\Delta \tau \approx 0~$ms (filled circles), 1~ms (squares), and 2~ms (triangles), respectively. In (a-d), horizontal dashed lines mark the quantum limit, below which Eq.~(\ref{skrelation}) is violated. Error bars include systematic and statistical errors. (e) Mean phonon population $\bar{N}_k$ (filled symbols) and pair-correlation amplitude $\Delta N_k$ (open symbols) extracted using the first minima and maxima identified in (a, circles), (b, squares), and (c, triangles), respectively. Blue (red) shaded areas mark the region where $\Delta N_k>\bar{N}_k$ ($\Delta N_k<\bar{N}_k$). Error bars represent statistical errors.}
\label{fig4}
\end{figure}

We now focus on identifying a key signature of non-classical correlations. To search for entanglement in the final phonon basis, we evaluate the squeezing parameter $C_k = \epsilon_k/\epsilon(k,g_f)$ at $g=g_f$ and plot the rescaled phonon spectra $\tilde{S}(k,\tilde{\tau})=S(k,\tilde{\tau})/C_k$, as shown in Figs.~\ref{fig4}(a-c). In this basis, the phonon spectra at momenta $k \gtrsim 1.5/\mu$m can be observed to oscillate above and below the rescaled quantum limit $\tilde{S} = 1$, showing signatures of two-mode squeezing and anti-squeezing as time evolves. The first minimum $\tilde{S}_\mathrm{min}$ identified at various momenta $k$ is plotted in Fig.~\ref{fig4}(d), in which we find that $\tilde{S}_\mathrm{min}$ violates the inequality Eq.~(\ref{skrelation}) in a wider range for the MI sample with $\Delta \tau \approx 1~$ms than it does for the samples without MI or with longer $\Delta \tau$. The strongest violation is in the range of $2.1/\mu$m$\lesssim k\lesssim 2.2/\mu$m and has average $\tilde{S}_\mathrm{min}\approx 0.77(7)<1$, compared to $\tilde{S}_\mathrm{min}\approx 0.84(8)$ without MI and $\tilde{S}_\mathrm{min}\approx 0.91(5)$ for $\Delta \tau \approx2~$ms.

To further interpret this result, we extract the mean phonon number $\bar{N}_k$ and the pair-correlation amplitude $\Delta N_k$ by using the first maximum $\tilde{S}_\mathrm{max}$ and minimum $\tilde{S}_\mathrm{min}$ identified in $\tilde{S}(k,\tilde{\tau})$ at each $k$ in Figs.~\ref{fig4}(a-c),
\begin{eqnarray}
\bar{N}_k &\approx& \frac{\tilde{S}_\mathrm{max} + \tilde{S}_\mathrm{min}}{2}-1 \nonumber \\
\Delta{N}_k &\approx& \frac{\tilde{S}_\mathrm{max} - \tilde{S}_\mathrm{min}}{2}. \label{dn}
\end{eqnarray}
As shown in Fig.~\ref{fig4}(e), both $\bar{N}_k$ and $\Delta{N}_k$ have comparably increased due to pair-production in MI samples of $\Delta \tau \neq 0$. Quantum entanglement appears to better prevail for $\Delta \tau \approx 1~$ms and at $k\gtrsim 1.5/\mu$m, where $\Delta N_k \gtrsim \bar{N}_k$. This may be understood as any excessive incoherent population $\bar{N}_k-\Delta{N}_k>0$ in our samples can be due partially to quasiparticle dissipation during the quench and partially to incoherent (thermal) phonons present in the initial state. The latter are better suppressed at $k> 1.5/\mu$m as $\epsilon(k,g_i) > k_B T \approx \hbar \times 1~$kHz.

In summary, we demonstrate that coherent quasiparticle pair-production can be initiated in atomic quantum gases quenched to an attractive interaction. We show that in situ density noise power spectrum provides a simple means to characterize coherence and quantum entanglement between quasiparticles of opposite momenta, by using a continuous-variable version of the Peres-Horodecki separability criterion. Strong pair-correlation signal has been observed and non-classical correlation has been identified, with two-mode squeezing $\tilde{S}_\mathrm{min}\approx 0.8<1$ below the quantum limit. Further reduction of initial incoherent phonon populations or of decoherence during pair-production processes may increase the non-classical signal in future experiments. Reaching $\tilde{S}<0.5$ could open up applications requiring Einstein–Podolsky–Rosen entangled quasiparticle pairs \cite{reid1988quantum,peise2015satisfying,fadel2018spatial,kunkel2018spatially,lange2018entanglement}. Our quench protocol and entanglement detection method may be extended to analyze entanglement distribution between non-causal regions before the interaction quench. Furthermore, in analogy to the discussion in Ref. \cite{klich2009quantum}, extending our analyses of two-mode quadrature variance to skewness \cite{armijo2010probing} and other higher-order correlation terms may provide necessary observables for probing entanglement entropy and transport in a quantum gas.

\bibliographystyle{apsrev4-2}
\bibliography{entanglement_arXiv.bbs}

\begin{acknowledgments}
We thank M. Kruczenski and Q. Zhou for discussions. This work is supported in part by DOE QuantISED program (Grant \# DE-SC0019202) and the W. F. Keck foundation. C.-A.C. and C.-L.H. acknowledge support by the NSF (Grant \# PHY-1848316).
\end{acknowledgments}

\newcommand\eq[1]{Eq.~(\ref{#1})}
\newcommand\eqs[2]{Eqs.~(\ref{#1}) and (\ref{#2})}
\newcommand\dpar[2]{\frac{\partial#1}{\partial#2}}
\def\be{\begin{equation}}
\def\ee{\end{equation}}
\def\ba{\begin{eqnarray*}}
\def\ea{\end{eqnarray*}}
\def\half{\frac{1}{2}}
\def\Tr{\mbox{Tr}}
\def\al{\hat{\alpha}}
\def\eps{\epsilon}
\def\ome{\omega}
\def\gam{\gamma}
\def\Gam{\Gamma}
\def\Ome{\Omega}
\def\lam{\lambda}
\def\la{\langle}
\def\ra{\rangle}
\def\x{{\bf x}}
\def\k{{\bf k}}
\def\tg{\tilde{g}}
\def\dt{\Delta\tau}
\def\dn{\delta n}
\def\nave{\bar{n}}
\def\snave{\sqrt{\bar{n}}}
\def\ortV{\frac{1}{\sqrt{V}}}
\def\nk{\hat{n}_{\bf k}}
\def\ak{\hat{a}_{\bf k}}
\def\ahat{\hat{a}}
\def\n{\hat{n}}
\def\b{\hat{b}}
\def\yh{\hat{y}}
\def\yl{\hat{\Theta}}
\def\bl{\hat{\beta}}
\def\A{\hat{A}}
\def\O{\hat{O}}
\def\c{\hat{c}}
\def\xh{\hat{x}}
\def\ph{\hat{p}}
\def\u{\hat{u}}
\def\v{\hat{v}}
\def\cA{{\cal A}}
\def\cB{{\cal B}}
\def\bin{\hat{b}_\mathrm{in}}
\def\tbin{\widetilde{b}_\mathrm{in}}
\def\Re{\mbox{Re}\,}
\def\Im{\mbox{Im}\,}
\def\tb{\widetilde{b}}
\def\cN{{\cal N}}
\renewcommand{\figurename}{Fig.}
\renewcommand{\thefigure}{SM\arabic{figure}}
\setcounter{figure}{0}
\renewcommand{\theequation}{S\arabic{equation}}

\onecolumngrid
\appendix*

\section*{Supplementary information}
\subsection*{Calibration procedure}
We calibrate our imaging system (effective numerical aperature N.A.$\approx0.35$) by using in situ density noise power spectrum measured from thermal gases confined in the box potential. We first load a 2D gas using the standard loading procedure described in \cite{chen2020observation}. We then quench the scattering length to a large and negative value at $a\approx-300~a_0$, where $a_0$ is the Bohr radius, and hold for $>100~$ms. During this time, the 2D sample suffers three-body recombination loss and heating, with $N\approx 7,000$ atoms remaining in the box and at a temperature up to $T=170\pm15~$nK independently measured in time-of-flight measurements. Following this heating procedure, thermal de Broglie wavelength $\lambda_\mathrm{dB}<0.5~\mu$m is much smaller than the image resolution. We record the density fluctuations by quenching the scattering length back to a nearly non-interacting value and perform in situ absorption imaging. We evaluate the density noise power spectrum of the high temperature normal gases, which can be used to accurately determine the modulation transfer function of our imaging system \cite{hung2011extracting} that terminates at $k\gtrsim 2.6/\mu$m. All the density noise power spectra presented in this letter are normalized by the atomic shot-noise calibrated modulation transfer function to remove systematic image aberrations.

\subsection*{Separability criterion}
In this section, we derive the separability criterion for quasiparticles of opposite momenta $\pm \k$ using the density observables discussed in the main text. We define $\al^{\dag}_{\k}$, $\al_{\k}$ as the quasiparticle creation and annihilation operators. The associated coordinate and momentum operators can then be written as
\ba
\xh_1 & = & \frac{1}{\sqrt{2}} (\al_\k + \al^\dagger_{\k}) \\
\xh_2 & = & \frac{1}{\sqrt{2}} (\al_{-\k} + \al^\dagger_{-\k}) \\
\ph_1 & = & \frac{i}{\sqrt{2}} (\al^\dagger_\k - \al_{\k}) \\
\ph_2 & = & \frac{i}{\sqrt{2}} (\al^\dagger_{-\k} - \al_{-\k}) .
\ea
These have canonical commutation relations (with the reduced Planck constant $\hbar = 1$). Next,
we consider variances of
\ba
\u & = & \xh_1 + \xh_2 = \yh_1 + \yh_2 \\
\v & = & \ph_1 - \ph_2 = i (\yh_2 - \yh_1),
\ea
where $\yh_1$ and $\yh_2$ are non-Hermitian but momentum-conserving operators
\ba
\yh_1 & = & \frac{1}{\sqrt{2}} (\al_\k + \al^\dagger_{-\k}) \\
\yh_2 & = & \frac{1}{\sqrt{2}} (\al_{-\k} + \al^\dagger_{\k}),
\ea
which commute with each other. By momentum conservation (in the states we consider), we have $\la \yh_1^2 \ra = \la \yh_2^2 \ra = 0$, and
\ba
\la \u^2 \ra & = & 2 \la \yh_1 \yh_2 \ra = \la \al_\k \al_{-\k} + \al_\k \al^\dagger_{\k} + \al^\dagger_{-\k} \al_{-\k}
+ \al^\dagger_{-\k} \al^\dagger_{\k} \ra \\
\la \v^2 \ra & = & 2 \la \yh_1 \yh_2 \ra = \la \u^2 \ra
\ea
The total variance is
\ba
\la \u^2 \ra + \la \v^2 \ra = 2 \la \al_\k \al_{-\k} + \al_\k \al^\dagger_{\k} + \al^\dagger_{-\k} \al_{-\k}
+ \al^\dagger_{-\k} \al^\dagger_{\k} \ra.
\ea
For a separable state, by the theorem proven in \cite{duan2000inseparability,simon2000peres}, the total variance must satisfy the following inequality
\begin{equation}
[\la \u^2 \ra + \la \v^2 \ra]_\mathrm{sep} \geq 2. \label{criterion}
\end{equation}
If we take a thermal state of $\al_{\pm \k}$, $\al^\dagger_{\pm \k}$ for an example, such as when a quenched superfluid has fully equilibrated, we have
\ba
[\la \u^2 \ra + \la \v^2 \ra]_\mathrm{therm} = 2 (\la \al_\k \al^\dagger_{\k} \ra + \la \al^\dagger_{-\k} \al_{-\k}\ra ) = 2(2 n_B + 1 ) >2,
\ea
satisfying the separability criterion. Here $n_B > 0$ is the Bose-Einstein distribution.

We now express the separability criterion Eq.~(\ref{criterion}) using density observables and associate it with the density noise power spectrum $S(\k)$. We begin by considering the density noise and phase of the order parameter in the momentum space. The density noise is calculated as $\dn(\x,t) = n(\x,t) - \nave$ at an arbitrary time $t$, where $\nave$ is the average density of the superfluid, assumed uniform and time-independent. We define the Fourier components $\n_\k$ and $\hat{\theta}_\k$ of the density noise and phase operators as follows:
\begin{eqnarray}
\delta\n(\x,t) &=& \frac{1}{V}\sum_{\k\neq 0} \n_\mathbf{k}(t) e^{i\k \cdot \x} \\
\hat{\theta}(\x,t) &=& \frac{1}{V} \sum_{\k\neq 0} \hat{\theta}_\k(t) e^{i\k \cdot \x} = \frac{1}{V} \sum_{\k\neq 0} \hat{\theta}^\dag_\k(t) e^{-i\k \cdot \x},
\end{eqnarray}
where $V$ is the volume of the gas. The commutation relation is $[\n(\x), -\hat{\theta}(\x') ] = i \delta(\x - \x')$, where $\delta(\x)$ is the Dirac delta-function. In Fourier components, it corresponds to $[\n_\k, -\hat{\theta}^\dag_{\k'}] = iV \delta_{\k,\k'}$, where $\delta_{\k,\k'}$ is the Kronecker delta.

In Fourier space, the density and phase operators are related to the single-particle
operators $\ahat^{(\dag)}_{\pm\k}$ by
\begin{eqnarray}
\hat{n}_\k & = & \sqrt{N} (\hat{a}_\k + \hat{a}_{-\k}^\dagger) \,, \label{nk} \\
\hat{\theta}_\k & = & \frac{iV}{2 \sqrt{N}} (\ahat_{-\k}^\dagger - \ahat_\k) \, , \label{tk}
\end{eqnarray}
where $N=\nave V$ is the total particle number, assumed to be largely accounted for
by ground state atoms. Through the Bogoliubov transformation, we can further express the same quantities in terms of the quasiparticle operators $\al^{(\dag)}_{\pm\k}$. We have
\begin{eqnarray}
\n_\k & = & \sqrt{NC_k} (\al_\k + \al^\dagger_{-\k}) \label{ank} \\
-\hat{\theta}^\dag_\k & = & \frac{iV}{2 \sqrt{N C_k} } (- \al_{-\k} + \al^\dagger_\k)
\end{eqnarray}
where $C_k = \epsilon_k/\epsilon(k)$, $\epsilon(k) = \sqrt{\epsilon_k^2 + 2 \frac{\hbar^2}{m}\bar{n} g \epsilon_k}$ is the Bogoliubov energy (at the interaction $g$ when the measurement takes place), $\eps_k = \frac{\hbar^2 |\k|^2}{2 m}$ is the single particle energy, and $m$ is the atomic mass.

Using Eq.~(\ref{ank}), we can relate the density noise power spectrum to the variances of two-mode quadrature
\begin{equation}
S(\k) = \frac{\la \n^\dag_\k \n_{\k} \ra}{N} = \frac{C_k}{2} \left[ \la u^2 \ra + \la v^2 \ra \right].
\end{equation}
Thus, in a separable state we would have
\begin{equation}
S(\k)= \frac{\la \n_\k^\dag \n_{\k} \ra_\mathrm{sep}}{N} \geq C_k,
\end{equation}
where we have applied the inequality (\ref{criterion}).

\subsection*{Evolution of quasiparticles coupled to a bath}
Both in the present study and former experiments \cite{hung2013cosmology,ranccon2013quench}, quasiparticles are observed to dissipate. In this section, we calculate the coherent evolution of density fluctuations following an interaction quench in the presence of dissipation. We consider an instantaneous quench after which the value of the interaction parameter $g$ can be either positive or negative. At the level of the quadratic Hamiltonian, we can consider each $(\k, -\k)$ pair of modes separately. The corresponding system Hamiltonian is
\be
\hat{H}(\k,-\k) =\frac{1}{V} \left[2 \nave\eps_k \hat{\theta}_{-\k} \hat{\theta}_\k
+ \frac{1}{\nave}\left( \frac{\eps_k}{2} + \nave\tilde{g} \right) \n_{-\k} \n_\k \right] \, ,
\label{Hsys}
\ee
where $\nave$ is the average dentity, which is assumed uniform and time-independent, and $\tg = \frac{\hbar^2}{m}g$. Note that $\hbar \hat{\theta}_\k$ is the canonical momentum conjugate to $\n_{-\k}$. In what follows, we set $\hbar = 1$.

The quadratic approximation neglects nonlinear effects, which can change population of a given pair of modes. For near-equilibrium states, one may consider taking nonlinear terms into account via a version of the Boltzmann equation, written in terms of Bogoliubov quasiparticles. For negative $g$, however, the uniform equilibrium is unstable, and the usual definition of a quasiparticle does not apply.
To qualitatively describe the effect of nonlinear terms in this case, we consider a model in which the loss (gain) of particles in a given momentum mode is due to Markovian quantum noise. The model is specified by stating which of the system operators the noise couples to. Here, we assume those to be the single particle operators
$\hat{a}_\k$, $\hat{a}^\dagger_\k$.

To compute the effect of the noise on the evolution of the system, we use the quantum Langevin equation (QLE) in the form presented in Ref.~\cite{gardiner1985input}. In this formalism, each noise channel is described by time-dependent ``in'' operators, subject to commutation relations
\be
[\bin(t) , \bin^\dagger(t') ] = \delta(t - t')
\label{com_rel}
\ee
(with other pairwise commutators equal to zero). If there is only one such channel, the evolution of any operator $\A$ characterizing the system is given by the QLE as follows \cite{gardiner1985input}:
\be
\dot{\A} = -i [\A, \hat{H}] - [\A, \O^\dagger] \left\{ \half \Gam \O + \sqrt{\Gam} \bin (t) \right\} + \left\{ \half \Gam \O^\dagger + \sqrt{\Gam} \bin^\dagger(t) \right\} [\A, \O] \, ,
\label{qle}
\ee
where $\O$ is the system operator that couples to the noise, and $\Gam$ is the strength (width) of that coupling.

When there are two or more independent noise channels, one can generalize \eq{qle} by defining separate $\bin$, $\O$, and $\Gam$ for each channel and summing up the corresponding noise terms in the equation. In the present case, we have two noise channels, which couple to the system operators
\ba
\O_1 = \ahat_{\k} \, , \hspace{3em} \O_2 = \ahat_{-\k} \, .
\ea
We call the corresponding noise operators $\bin$ and $\tbin$, respectively. In what follows, we assume that the two widths are equal: $\Gam_1 = \Gam_2 = \Gam$. We allow the width to depend on $k = |\k|$, even though for brevity we do not attach a subscript $k$ to $\Gam$.

We now consider evolution of the system with the Hamiltonian (\ref{Hsys}) after an instantaneous quench of the coupling $\tg$ from a positive initial value $\tg_i$ to a final value $\tg_f$; the latter can be positive or negative. It is convenient to define, instead of the phase component $\hat{\theta}_\k$, a related variable $\yl_\k \equiv -2 \nave \hat{\theta}_\k$. Then, the QLEs for $\yl_\k$ and $\n_\k$ can be written in the matrix form as
\be
\left( \begin{array}{c} \dot{\yl}_\k \\ \dot{\n}_\k \end{array} \right)
= \left( \begin{array}{cc} - \half \Gam & \eps_k + 2 \nave \tg \\
- \eps_k & -\half \Gam \end{array} \right)
\left( \begin{array}{c} \yl_\k \\ \n_\k \end{array} \right)
- \sqrt{N\Gam} \left( \begin{array}{c} i [\bin(t) - \tbin^\dagger(t) ] \\
\bin(t) + \tbin^\dagger(t)
 \end{array} \right) .
\label{qle_mat}
\ee
We search for solution as an expansion in the eigenvectors of the matrix in (\ref{qle}), that is,
\be
\left( \begin{array}{c} \yl_\k \\ \n_\k \end{array} \right)
= \c_1(t) \left( \begin{array}{c} - i \frac{\ome_k }{ \eps_k} \\ 1 \end{array} \right)
+ \c_2(t) \left( \begin{array}{c} i \frac{\ome_k }{ \eps_k} \\ 1 \end{array} \right) \, ,
\label{sol}
\ee
where
\[
\ome_k=\sqrt{\eps_k^2 + 2 \nave \tg_f \eps_k}
\]
is Bogoliubov's quasiparticle frequency, and $\tg_f$ is the coupling constant after the quench. For negative $\tg_f$, there is an instability band, corresponding to those $k$ for which the argument of the square root is negative. For these $k$, $\ome_k$ is imaginary; by convention we choose it to have a negative imaginary part.

Equations satisfied by $\c_{1,2}$ are
\ba
\dot{\c}_1 & = & \lam_+ \c_1 - \sqrt{N\Gam} \zeta_+(t) \, , \\
\dot{\c}_2 & = & \lam_- \c_2 - \sqrt{N\Gam} \zeta_-(t) \, ,
\ea
where
\ba
\lam_{\pm} = - \half \Gam \pm i \ome_k
\ea
are the eigenvalues of the evolution matrix, and
\be
\hat{\zeta}_\pm(t) = \half \left( 1 \mp \frac{\eps_k}{\ome_k} \right) \bin(t)
+ \half \left( 1 \pm \frac{\eps_k}{\ome_k} \right) \tbin^\dagger(t) \, .
\label{zeta}
\ee
According to \eq{sol}, $\n_\k(t) = \c_1(t) + \c_2(t)$. The general solution for it is
\be
\n_\k(t) = \c_1(0) e^{\lam_+ t} + \c_2(0) e^{\lam_- t} - \sqrt{N\Gam}
\int_0^t \left[ \hat{\zeta}_+(t') e^{\lam_+ ( t-t')} + \hat{\zeta}_-(t') e^{\lam_- ( t-t')} \right] dt' \, .
\label{nk_sol}
\ee
The operator constants $\c_1(0)$ and $\c_2(0)$ are determined from the initial conditions. Let $\bl^\dagger_\k$, $\bl_\k$ be the creation and annihilation operators of Bogoliubov's quasiparticles before the quench. Then, at $t\to 0^-$,
\ba
\yl_\k(0^-) & = & iV\sqrt{\frac{ N}{C'_k}} (\bl_\k - \bl_\k^\dagger) \, , \\
\n_\k(0^-) & = & \sqrt{NC'_k} (\bl_\k + \bl_{-\k}^\dagger) \, ,
\ea
where $C'_k \equiv \eps_k / \Ome_k$, and
\ba
\Ome_k = \sqrt{\eps_k^2 + 2 \nave \tg_i \eps_k}
\ea
is the initial-state quasiparticle frequency. Equating these $\yl_\k$, $\n_\k$ to their $t\to 0^+$ limits, obtained from \eq{sol}, we find
\begin{equation*}
\c_{1,2}(0) = \frac{\sqrt{NC'_k }}{2} \left[ \bl_\k \left( 1 \mp \frac{\Ome_k}{\ome_k} \right) + \bl_{-\k}^\dagger \left( 1 \pm \frac{\Ome_k}{\ome_k} \right) \right] \, ,
\end{equation*}
where the upper signs are for $\c_1$, and the lower for $\c_2$.

For computation of the power spectrum of density fluctuations, we will need both $\n_\k$ and $\n_{-\k}$. The former is obtained by substituting the above expressions for $\c_{1,2}(0)$ into \eq{nk_sol}. For the latter, we need to exchange $\k$ with $-\k$ in the expressions for $\c_{1,2}(0)$. In addition, recalling that untilded noise operators refer to mode number $\k$, and tilded ones to $-\k$, we need to exchange the tilded and untilded operators in \eq{zeta}.

The quasiparticle operators $\bl$, $\bl^\dagger$ can in principle depend on noise, but only on that part of it that is represented by $\bin(t)$, $\tbin(t)$ with $t < 0$. None of those appear in the solution (\ref{nk_sol}). Thus, we can consider $\c_{1,2}(0)$ on the one hand, and the noise operators appearing in \eq{nk_sol} on the other to be distributed independently. Further, we assume that all the one-point functions, i.e., $\la \bl \ra$, $\la \bin \ra$, $\la \tbin \ra$, of these operators are zero. As a result, the power spectrum will consist of two separate parts,
\be
S(k)=\frac{\la \n_{-\k} \n_{\k} \ra}{N} = \frac{\la \n_{-\k} \n_\k \ra_\mathrm{coh}}{N} + \frac{\la \n_{-\k} \n_\k \ra_\mathrm{inc}}{N} \, ,
\label{pws}
\ee
which we refer to as coherent and incoherent. The former corresponds to the first two terms in (\ref{nk_sol}) and the latter to the last term, due entirely to the noise. Outside the instability band, where $\ome_k$ is real, the coherent part decays at large times, while the incoherent part goes to a constant value. For $k$ in the instability band, both parts can be amplified; that is the case when $|\mathrm{Im}[\ome_k]| > \Gam/ 2$, where $\mathrm{Im}[.]$ denotes imaginary part.

Finally, we assume that, among the two-point functions of the aforementioned operators, the only nonzero ones are
\begin{equation*}
\la \bl^\dagger_\k \bl_\k \ra = \cN_k \, , \hspace{3em}
\la \bl_\k \bl^\dagger_\k \ra = \cN_k + 1 \, ,
\end{equation*}
\begin{equation*}
\la \bin(t) \bin^\dagger(t') \ra = \la \tbin(t) \tbin^\dagger(t') \ra= \delta(t-t') \, .
\end{equation*}
The latter is a simplifying assumption to the effect that the noise is in vacuum. One may reasonably
hope it to work for sufficiently short evolution times. With these assumptions, we obtain
\begin{equation*}
\frac{\la \n_{-\k} \n_\k \ra_\mathrm{coh} }{N}= e^{-\Gam t} S_\mathrm{coh}(k)= e^{-\Gam t} S_i(k)
\left[ 1 + \left( \frac{\Ome_k^2}{\ome_k^2} - 1 \right) \sin^2 \ome_k t \right],
\end{equation*}
where $S_i(k)= C_k' (2 \cN_k + 1) $ is the initial density noise power spectrum and
\begin{align*}
&\frac{\la \n_{-\k} \n_\k \ra_\mathrm{inc} }{N} = S_\mathrm{inc}(k) \\
&=\half \left\{ \left(1 - \frac{\eps_k^2}{\ome_k^2} \right) \frac{\Gam}{\Gam^2 + 4 \ome_k^2} \left[ \Gam \left( 1 - e^{-\Gam t} \cos 2 \ome_k t \right) + 2 \ome_k e^{-\Gam t} \sin 2 \ome_k t \right] + \left( 1 + \frac{\eps_k^2}{\ome_k^2} \right) \left( 1 - e^{-\Gam t}\right) \right\}
\end{align*}

We compare the above theory calculation with the density noise power spectra measured after we quench the interaction strength to $g=g_\mathrm{MI}$ and hold for a time $t=\dt$, as shown in Fig.~\ref{fig2}. We assume a $k$-dependent coupling rate $\Gamma_k = \eta \epsilon_k$ and set $\eta$ as a fit parameter. We then use the measured initial spectrum $S_i(k)$ and all other experimentally measured quantities to evaluate the power spectrum $S(k) = e^{-\Gam_k \dt} S_\mathrm{coh}(k) + S_\mathrm{inc}(k)$. As shown in Fig.~SM1, we find a reasonable fit with $\eta \approx 0.54(2)$ for samples with a hold time of $\Delta \tau \approx 2~$ms.

\begin{figure}[t]
\includegraphics[width=0.4\columnwidth]{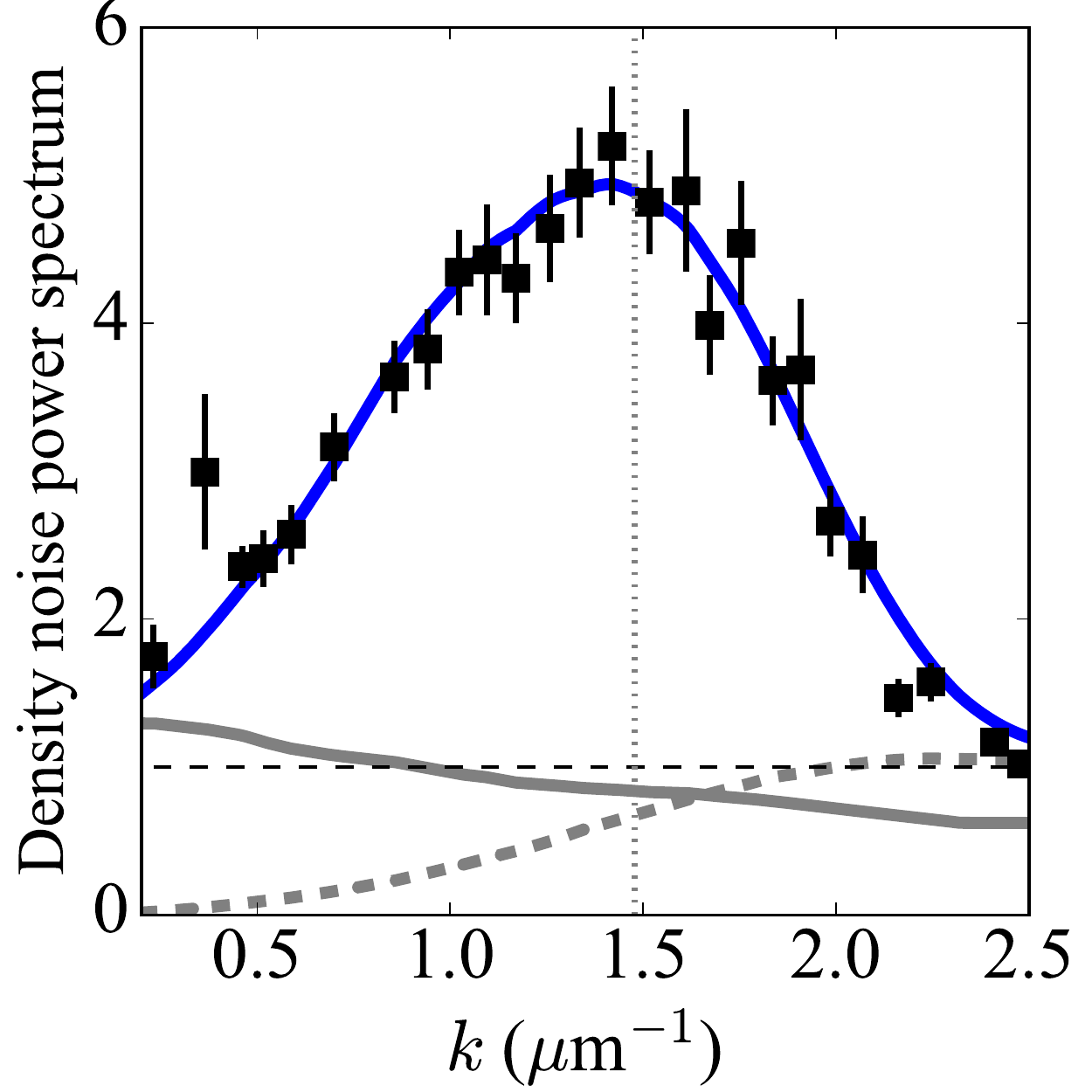}
\caption{Fitting the measured density noise power spectra (symbols) right after the MI period with $\Delta\tau \approx 2$~ms. Blue curve shows the fit, calculated based on the measured initial spectrum (smoothed gray curve). Dashed curve shows the fitted incoherent contribution $S_\mathrm{inc}(k)$. Vertical dotted line marks the wavenumber $k_c$, below (above) which quasiparticles are expected to be unstable (stable).}
\label{figSM}
\end{figure}

\end{document}